\documentclass[aps,pre,twocolumn,amsmath,amsfonts,showpacs]{revtex4}
\newcommand{\bec}[1]{\mbox{\boldmath $ #1$}}
\usepackage{graphicx}
\begin{document}
\title{Effect of large-scale coherent structures on turbulent convection}
\author{M. Bukai}
\email{mbukai@gmail.com}
\author{A. Eidelman}
\email{eidel@bgu.ac.il}
\author{T. Elperin}
\email{elperin@bgu.ac.il}
\homepage{http://www.bgu.ac.il/me/stuff/tov}
\author{N. Kleeorin}
\email{nat@bgu.ac.il}
\author{I. Rogachevskii}
\email{gary@bgu.ac.il}
\homepage{http://www.bgu.ac.il/~gary}
\author{I. Sapir-Katiraie}
\email{katiraie@bgu.ac.il}
\affiliation{The Pearlstone Center for Aeronautical Engineering
Studies, Department of Mechanical Engineering,
Ben-Gurion University of the Negev, P.O.Box 653, Beer-Sheva 84105,  Israel}
\date{\today}
\begin{abstract}
We study an effect of large-scale coherent structures on global properties of turbulent convection in laboratory experiments in air flow in a  rectangular chamber with aspect ratios $A \approx 2$ and $A\approx 4$ (with the Rayleigh numbers varying in the range from $5 \times 10^6$ to $10^8$). The large-scale coherent structures comprise the one-cell
and two-cell flow patterns. We found that a main contribution to the turbulence kinetic energy production in turbulent convection with large-scale coherent structures is due to the non-uniform large-scale motions. Turbulence in large Rayleigh number convection with coherent structures is produced by shear, rather than by buoyancy. We determined the scalings of global parameters (e.g., the production and dissipation of turbulent kinetic energy, the turbulent velocity and integral turbulent scale, the large-scale shear, etc.) of turbulent convection versus the temperature difference between the bottom and the top walls of the chamber. These scalings are in an agreement with our theoretical predictions. We demonstrated that the degree of inhomogeneity of the turbulent convection with large-scale coherent structures is small.
\end{abstract}

\pacs{47.27.te, 47.27.-i}

\maketitle

\section{Introduction}

Coherent structures in a turbulent convection at large Rayleigh numbers have been observed in the atmospheric turbulent convection
\cite{EB93,AZ96,LS80,H84,ET85,W87,H88,SS89,Z91,WK96,ZGH98,BR99,YKH02,EKRZ06} (so-called ''semi-organized structures''), in numerous laboratory experiments in the Rayleigh-B\'{e}nard apparatus
\cite{KH81,SWL89,SI94,CCL96,K01,NSS01,SB02,NS03,QT01,BKT03,XSZ03,LLT05,XL04,SQ04,BNA05,VTH06,EEKR06,RPT06,BA06,PINT07,XX07,HT09}
(so-called ''wind'' or ''mean wind'') and in direct numerical simulations \cite{KER01,HTB03,PHP04,RJH05}. Characteristic spatial and time scales of the coherent structures in a turbulent convection are larger than turbulent scales.

In spite of numerous theoretical, experimental and
numerical studies of coherent structures in turbulent convection,
their origin is still a subject of numerous discussions.
In particular, there are two points of view on the origin of coherent structures in turbulent convection \cite{HTB03}. According to one
point of view, the rolls which develop at low Rayleigh numbers near
the onset of convection continually increase their size as Rayleigh
number is increased and continue to exist in an average sense at
even the highest Rayleigh numbers reached in the experiments
\cite{F76}. Another opinion consists in that coherent structures are genuine high Rayleigh number effect \cite{KH81}.

A new mean-field theory of formation of coherent structures in turbulent convection has been suggested in \cite{EKRZ02,EKRZ06,EGKR06}. According to this theory a redistribution of the turbulent heat flux due to non-uniform large-scale motions plays a crucial role in the formation of the large-scale coherent structures in turbulent convection. A convective-wind instability in the shear-free turbulent
convection causes the large-scale motions in
the form of cells. In the sheared convection, the large-scale
instability results in an excitation of convective-shear waves, and the
dominant coherent structures in the sheared convection are rolls \cite{EKRZ02,EKRZ06,EGKR06}.

It was suggested on the base of laboratory experiments and numerical simulations performed in \cite{BKT03} that the coherent structures in  the Rayleigh-B\'{e}nard turbulent convection are not driven by the turbulent Reynolds stresses, associated with the tilting plumes at the upper and the lower horizontal walls. The numerical results in \cite{BKT03} show that once the mean flow is established, the temperature of the fluid is larger at one side wall and smaller at the other side, and the mean flow is driven by the mean buoyant force at the side walls. This is also in agreement with the experimental studies in \cite{XSZ03,VTH06}.

There is an opinion that buoyancy plays a crucial role in production of turbulent convection. For example, in the mixing length theory of astrophysical turbulent convection the following estimate for the turbulent kinetic energy is often used \cite{PR82,RU89,S89}:
\begin{eqnarray}
\langle {\bf u}^2 \rangle/2 = g \, \tau \, \langle u_z \, s \rangle ,
\label{A1}
\end{eqnarray}
where $\langle u_z s \rangle$ is the vertical turbulent flux of entropy, ${\bf u}$ and $s$ are fluctuations of fluid velocity and entropy, ${\bf g}$ is the acceleration of gravity, $\tau$ is the characteristic correlation time of turbulent velocity field and angular brackets denote ensemble averaging. Equation~(\ref{A1}) implies that the vertical turbulent flux of entropy plays a role of a stirring force for the turbulent convection.

However, this estimate can be valid only in the absence of large-scale coherent structures. Measurements of the mean temperature distributions and heat fluxes in laboratory experiments in the Rayleigh-B\'{e}nard turbulent convection (see, e.g., \cite{SQ04,EEKR06,XLZ02}) show that a thermal structure inside the large-scale circulation is inhomogeneous and anisotropic. The hot thermal plumes accumulate at one side of the large-scale circulation, and cold plumes concentrate at the opposite side of the large-scale circulation. These measurements demonstrate that the vertical turbulent heat flux inside the large-scale circulation is very small in spite of large temperature difference $\Delta T$ between the bottom and the top  walls of the chamber.  On the other hand, the horizontal turbulent heat flux inside the large-scale circulation is larger than the vertical turbulent heat flux in spite of the absence of the imposed temperature difference $(\Delta T)_y$ between the side walls of the chamber (see, e.g., \cite{SQ04}). The mean velocity field inside the large-scale circulation is strongly non-uniform and turbulence can be also produced by mean velocity shear. Therefore, one may ask the question about the origin of turbulence inside the coherent structures, i.e., is it shear-produced turbulence or buoyancy-produced turbulence?

The goal of this study is to investigate experimentally an effect of large-scale coherent structures on global properties of turbulent convection. In particular, we address the following issues: (i) the origin of production in turbulent convection with large-scale coherent structures (a shear-produced or buoyancy-produced turbulence); (ii) the scalings of global parameters (production and dissipation of turbulent kinetic energy, turbulent velocity and integral turbulent scale, the large-scale shear, etc.) versus the temperature difference $\Delta T$ between the bottom and the top  walls of the chamber; (iii) the degree of inhomogeneity of the turbulent convection with large-scale coherent structures.

The paper is organized as follows. Section II
describes the experimental set-up for a laboratory study of the coherent structures. The experimental results and their detailed analysis are
presented in Section III. Finally, conclusions are drawn in Section
IV.

\section{Experimental set-up}

\begin{figure}
\vspace*{1mm}
\centering
\includegraphics[width=8cm]{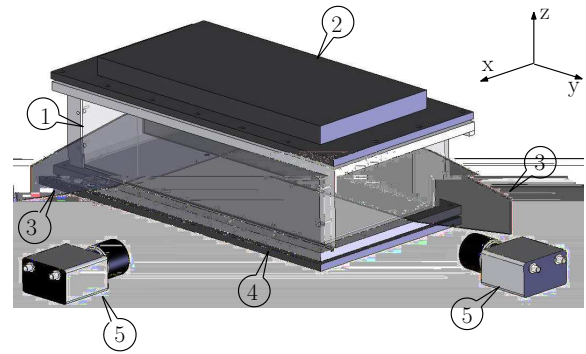}
\caption{\label{Fig1} Experimental set-up: (1) - walls of the
chamber; (2) - cooled top with a heat exchanger; (3) - laser light sheets; (4) - heated bottom with a heat exchanger; (5) - two digital CCD cameras.}
\end{figure}

The experiments were conducted in two rectangular chambers with dimensions $26 \times 58 \times 26$ cm$^3$ and $26 \times 58 \times 13$ cm$^3$. Hereafter, we use the following system of coordinates: $z$ is the vertical axis, the $y$-axis is directed along the longest wall (see Fig.~\ref{Fig1}). The side walls of the chambers are made of
transparent Perspex with the thickness of $10$ mm.
A number of experiments have been conducted with different additional thermal insulation of the side walls of the chamber in order to study
whether a heat flux through the side walls affects the turbulent
convective pattern. First, the side walls of the chamber were
insulated with Styrofoam plates with low thermal conductivity
($\kappa \sim 0.033$ W/mK) and with the thickness of 30 mm. Two
of the side plates were removed for a short time when the images of
the flow were recorded. In the next series of experiments we
installed additional Perspex plates with a thickness of 6 mm and
with an air gap of 2 mm between these plates and the outside walls
of the chamber. Finally, we performed experiments where these two
types of thermal insulation were used simultaneously. All these experiments have shown that the heat flux through the side walls does not affect the turbulent convective pattern.

A vertical mean temperature gradient in the turbulent air flow was
formed by attaching two aluminium heat exchangers to the bottom and
top walls of the test section (a heated bottom and a cooled top wall
of the chamber). The thickness of the aluminium plates is 2.5 cm.
The top plate is a bottom wall of the tank with cooling water.
Temperature of water circulating through the tank and the chiller is
kept constant within $0.1$ K. Cold water is pumped into the chamber
through two inlets and flows out through two outlets located at the
side wall of the chamber. The bottom plate is attached to the
electrical heater that provides constant and uniform heating. The
voltage of a stable power supply applied to the heater varies up to 200 V. The power of the heater varies up to 300 W. The temperatures of the conducting plates were measured with four thermocouples attached at the surface of each plate.

The temperature difference  between the top and bottom
plates, $\Delta T$,  varies in the range from 20 K to 80 K depending
on the power of the heater (i.e., the global Rayleigh number based on
molecular transport coefficients, was changed within the range of ${\rm Ra} = (0.4 - 1) \times 10^8$ for the aspect ratio of the chamber $A =H_y/H_z = 2.23$ and ${\rm Ra} = (0.5 - 1.2) \times 10^7$ for $A= 4.46$). Here $H_y$ and $H_z$ are the sizes of the chamber along $y$-axis and $z$-axis, respectively. The temperature in the probed region was measured with a specially designed temperature probe with twelve sensitive thermocouples having the diameter 0.125 mm. This allows us to obtain the detailed mean temperature distributions inside the coherent structures.

The velocity field was measured using a digital Particle Image Velocimetry (PIV), see \cite{RWK98}. In the experiments we used LaVision Flow Master III system. A double-pulsed light sheet was provided
by a Nd-YAG laser (Continuum Surelite $ 2 \times 170$ mJ). The light
sheet optics includes spherical and cylindrical Galilei telescopes
with tuneable divergence and adjustable focus length. We used two
progressive-scan 12 bit digital CCD cameras (with pixel size $7.4 \,
\mu$m $\times 7.4 \, \mu$m and $2048 \times 2048$ pixels) with a
dual-frame-technique for cross-correlation processing of captured
images. A programmable Timing Unit (PC interface card) generated
sequences of pulses to control the laser, camera and data
acquisition rate. The software package LaVision DaVis 7 was applied
to control all hardware components and for 32 bit image acquisition
and visualization. This software package comprises PIV software for
calculating the flow fields using cross-correlation analysis. The velocity field was measured in the two perpendicular planes (in five $yz$ planes and in ten $xz$ planes, so that the distance between these planes is 5 cm, see Fig.~\ref{Fig1}).

An incense smoke with sub-micron particles as a tracer was used for
the PIV measurements. Smoke was produced by high temperature
sublimation of solid incense particles. Analysis of smoke particles
using a microscope (Nikon, Epiphot with an amplification of 560) and
a PM-300 portable laser particulate analyzer showed that these
particles have an approximately spherical shape and that their mean
diameter is of the order of $0.7 \mu$m. The probability density
function of the particle size measured with the PM300 particulate
analyzer was independent of the location in the flow for incense
particle size of $0.5-1 \, \mu $m.

Series of 260 pairs of images acquired with a frequency of 1 Hz were
stored for calculating the velocity maps and for ensemble and
spatial averaging of turbulence characteristics. The center of the
probed flow region coincides with the center of the chamber. We
measured the velocity field in different flow areas, e.g., in the chamber with the aspect ratio $A\approx 2$, the probed flow area is $492 \times 212.5$ mm$^2$ with a spatial resolution of $302 \; \mu$m~/~pixel in $yz$ plane, and the probed flow area is $237 \times 211$ mm$^2$ with a spatial resolution of $264 \; \mu$m~/~pixel in $xz$ plane. Similarly, in the chamber with the aspect ratio $A\approx 4$, the probed flow area is $547 \times 127$ mm$^2$ with a spatial resolution of $278  \; \mu$m~/~pixel in $yz$ plane, and the probed flow area is $256 \times 125$ mm$^2$ with a spatial resolution of $130  \; \mu$m~/~pixel in $xz$ plane. These regions were analyzed with interrogation window of $32 \times 32$ pixels in the chamber with the aspect ratio $A\approx 2$ and with interrogation window of $24 \times 24$ pixels in the chamber with the aspect ratio $A\approx 4$. A velocity vector was determined in every interrogation window, allowing us to construct a velocity map comprising up to $3655$ vectors.

Mean and r.m.s. velocities, two-point correlation functions and
integral scales of turbulence were determined from the measured
velocity fields. The mean and r.m.s. velocities for each point of
the velocity map (up to 3280 points) were determined by averaging over 260 independent maps, and then over 3280 points. The two-point longitudinal correlation functions of the velocity field were determined for the central part of the velocity map by averaging over 260 independent velocity maps. In the experiments we evaluated the variability between the fist and the last 20 frames. Since no variability was found, these tests showed that 260 image pairs contain enough data to obtain reliable statistical estimates. The characteristic turbulence time in the experiments $\tau_z = 0.8 - 1.2$ seconds, while the characteristic time period for the large-scale circulatory flow is by one order of magnitude larger than $\tau_z$. These two characteristic times are much smaller than the time during which the velocity fields are measured.

An integral scale $\ell$ of turbulence was determined from the two-point correlation functions of the velocity field. These measurements were repeated for various temperature differences between the bottom and the top walls of the chamber. The size of the probed region did not affect our results. Similar experimental set-up and data processing procedure were used in experimental study of hysteresis phenomenon in turbulent convection in \cite{EEKR06} and in \cite{BEE04,EE04,EEKR06C,EEKR06A} for investigating a new phenomenon of turbulent thermal diffusion \cite{EKR96,EKRS00}.

The maximum tracer particle displacement in the experiment was of
the order of $1/4$ of the interrogation window.
The average displacement of tracer particles was of the order of
$2.5$ pixels. The average accuracy of the velocity
measurements was of the order of $4 \%$ for the accuracy of the
correlation peak detection in the interrogation window of the order
of $0.1$ pixels (see, e.g., \cite{AD91,RWK98,W00}).

In order to measure the vertical turbulent heat flux $F_z =  \langle u_z \theta \rangle$, the velocity $u_z$ and temperature $\theta$ fluctuations are determined using the temperature measurements by a thermocouple simultaneously with the velocity measurements by the PIV system. A time constant of the thermocouple was evaluated experimentally as $75$ ms that is close to the Kolmogorov's time of turbulence. The temperature is acquired in a time interval between two laser pulses used for the acquisition of every velocity vector map. During this time interval we obtain the data set consisting of about 100 measured temperatures. In order to decrease a noise level we use the mean value of this data set. Simultaneous temperature and velocity measurements should be conducted at the same point in order to determine the vertical turbulent heat flux $F_z$. Since the distance between the light sheet and the thermocouple is $3$ mm, we have to use a correction factor $1/0.95$ to a measured value of the heat flux. This correction factor is obtained using the measured two-point correlation function of the velocity field. Indeed, the magnitude of the longitudinal correlation function of the velocity field at the distance $3$ mm, is 0.95.

\section{Production and dissipation in turbulent convection}

\begin{figure}
\vspace*{1mm}
\centering
\includegraphics[width=7cm]{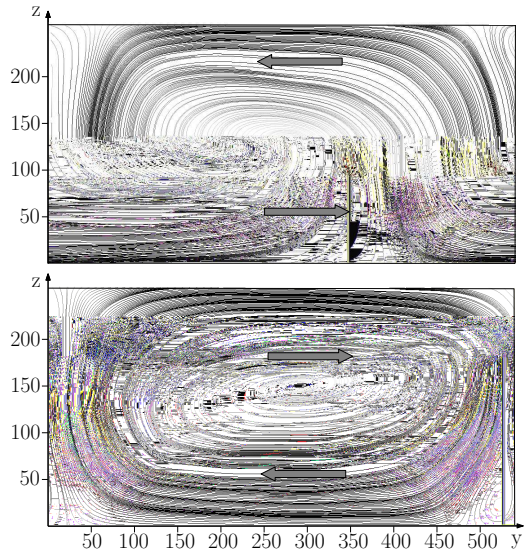}
\caption{\label{Fig2} Mean flow patterns obtained in the experiments with turbulent convection in the chamber with $A\approx 2$: one-cell flow patterns with the counter clockwise motions at the temperature difference $\Delta T = 20$ K (upper panel); one-cell flow patterns with the clockwise motions at the temperature difference $\Delta T = 35$ K (lower panel). Coordinates $y$ and $z$ are measured in mm.}
\end{figure}

\begin{figure}
\vspace*{1mm}
\centering
\includegraphics[width=8cm]{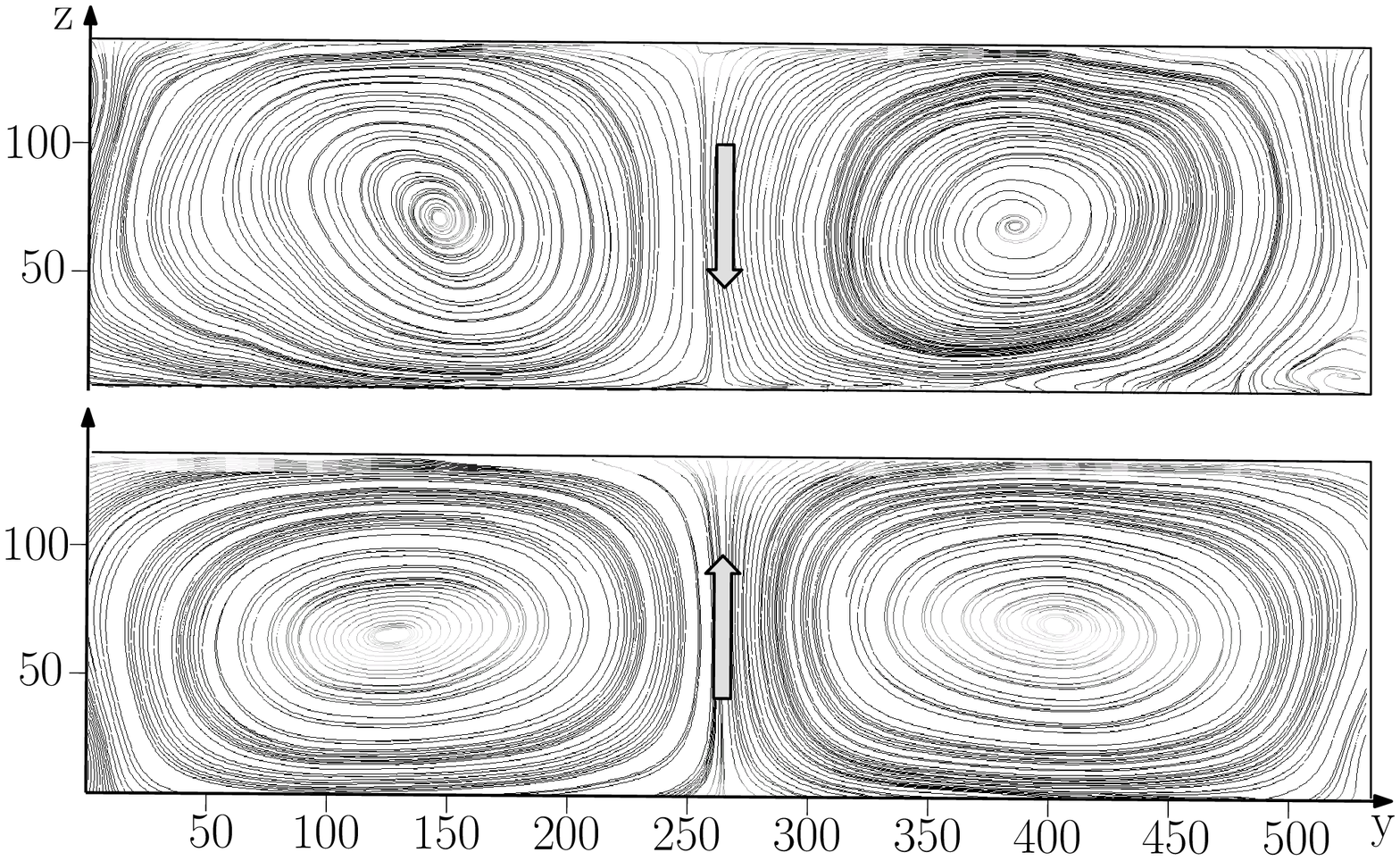}
\caption{\label{Fig3} Mean flow patterns obtained in the experiments with turbulent convection in the chamber with $A\approx 4$: two-cell flow patterns with  the downward motions in the central region of the chamber between two cells at the temperature difference $\Delta T = 20$ K (upper panel); two-cell flow patterns with  the upward motions in the central region of the chamber between two cells at the temperature difference $\Delta T = 33$ K (lower panel). Coordinates $y$ and $z$ are measured in mm.}
\end{figure}

Our velocity measurements show that the observed large-scale coherent structures comprise the one-cell and two-cell flow patterns. For instance, in Figs.~\ref{Fig2} and ~\ref{Fig3} we show the mean flow patterns obtained in the experiments with turbulent convection in the chambers with $A\approx 2$ and $A\approx 4$. The prevailing observed flow pattern in the experiments in the chamber with the aspect ratio $A \approx 2$ is one-cell flow structure (see Fig.~\ref{Fig2}), while the prevailing observed flow pattern in the experiments in the chamber with the aspect ratio $A \approx 4$ is two-cell flow structure (see Fig.~\ref{Fig3}). The prevailing observed flow pattern is the flow pattern which is observed in the largest range of the Rayleigh numbers. This feature is in agreement with the theoretical predictions made in \cite{EGKR06}. In particular, the threshold required for the excitation of the large-scale instability that causes formation of large-scale coherent structures, is minimum when the ratio of the horizontal size to the vertical size of the large-scale circulation is approximately 2 \cite{EGKR06}. This value corresponds to the one-cell flow pattern in the experiments with $A \approx 2$, and to the two-cell flow pattern in the experiments with $A \approx 4$. Note also that depending on the temperature difference $\Delta T$ between the bottom and the top  walls of the chamber, we observe in the chamber with $A\approx 4$ the two-cell flow patterns with the downward (for $\Delta T < 26$ K) or upward (for $\Delta T > 26$ K) motions in the central region of the chamber between two cells (see Fig.~\ref{Fig3}).

Note that the large-scale circulations in a turbulent convection have been previously experimentally studied in \cite{VTH06} in a Rayleigh-B\'{e}nard cell with the aspect ratio $A=4$ (with sizes $60 \times 60 \times 15.5$ cm$^3$). Experiments conducted in this cell filled with water (with Prandtl number Pr$=5.5$) and at a global Rayleigh number Ra$=5.9  \times 10^8$ (corresponding to a temperature difference of $6.3$ K), revealed that the mean flow field measured with PIV comprises two rolls. This result was confirmed in our experiments for the similar aspect ratio of the chamber $A=4.45$, but in air flow (with Prandtl number Pr$=0.7$) and with different horizontal cross section of the cell: rectangle in our experiments and square in the experiments reported in \cite{VTH06}.

\begin{figure}
\vspace*{1mm}
\centering
\includegraphics[width=8cm]{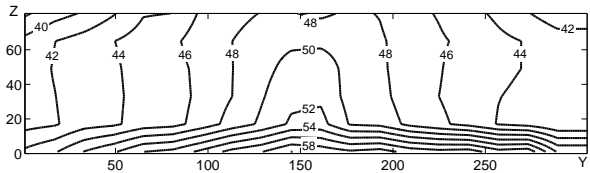}
\caption{\label{Fig4} Mean temperature field obtained in the experiments with turbulent convection in the chamber with $A\approx 4$ in the $yz$ plane for the two-cell flow patterns with the upward motions in the central region of the chamber between two cells. The mean temperature field is shown in degrees of C. Coordinates $y$ and $z$ are measured in mm.}
\end{figure}

The mean temperature measurements show that the thermal structure inside the large-scale circulation is inhomogeneous and anisotropic. For instance, in Fig.~\ref{Fig4} we show the isolines of the mean temperature field obtained in the experiments with turbulent convection in the chamber with $A\approx 4$ in the $yz$ plane for the two-cell flow patterns with the upward motions in the central region of the chamber between two cells. The hot thermal plumes accumulate at one side of the large-scale circulations (in the central part of the flow), and cold plumes concentrate at the opposite side of the large-scale circulations (in the periphery part of the flow near side walls). This fact is in agreement with the temperature measurements performed in \cite{SQ04}.

Now let us discuss the origin of the production and dissipation in turbulent convection with large-scale coherent structures. Equation for the evolution of the turbulent kinetic energy
$E_K=\langle {\bf u}^2 \rangle/2$ reads:
\begin{eqnarray}
{DE_K \over Dt} + {\rm div} \, {\bf \Phi}_{K} &=& -\langle u_i \, u_j \rangle \, \nabla_j U_i + \beta \, F_z - \varepsilon \;,
\label{A2}
\end{eqnarray}
where $D / Dt = \partial / \partial t + {\bf U} {\bf \cdot} \bec{\nabla} $, $\; {\bf u}$ are the fluctuation of the fluid velocity, ${\bf U}$ is the mean velocity that describes coherent structure, ${\bf \Phi}_{K}$ is the term that includes the third-order moments, $\beta=g/T_\ast$  is the buoyancy parameter and $T_\ast$ is a reference value of the mean absolute temperature, ${\bf g}$ is the acceleration of gravity, $F_i = \langle u_i \theta \rangle$ is the heat flux, $\theta$ are the temperature fluctuations, $\varepsilon = C_\varepsilon \, E_K/\tau$ is the dissipation rate of the turbulent kinetic energy. While the magnitude of the constant $C_\varepsilon$, based on several experiments that were summarized in \cite{SR84} and on the data obtained in \cite{BA05}, exhibits a relatively large scatter in the range of Reynolds numbers ${\rm Re}_\lambda < 50$, on the average the value of the constant $C_\varepsilon$  is closed to 1. Here ${\rm Re}_\lambda$ is the Reynolds number based on the Taylor microscale. The value of the constant $C_\varepsilon=1$ was obtained for the first time in \cite{B53} based on the experimental study \cite{BT48} for $14 < {\rm Re}_\lambda < 41$.

In order to determine experimentally the turbulent kinetic energy dissipation rate $\varepsilon$, we take into account a small anisotropy of the turbulent velocity field. In particular, we determine $\varepsilon$ as follows:
\begin{eqnarray}
\varepsilon = {C_\varepsilon \over 2} \Big({u_x^3 \over \ell_x}+{u_y^3 \over \ell_y}+{u_z^3 \over \ell_z}\Big )\;,
\label{A20}
\end{eqnarray}
where $u_x$, $\, u_y$ and $u_z$ are the components of the turbulent velocity field (rms), $\ell_x$, $\, \ell_y$ and $\ell_z$ are the integral turbulent length scales along $x$, $y$ and $z$ axis, and $C_\varepsilon=1$. In our experiments we determine the turbulent length scales and three components of turbulent velocity fluctuations in the range of ${\rm Re}_\lambda$  from 30 to 55, and for the temperature difference range $\Delta T$ between the bottom and the top walls of the chamber from 19 K to 80 K. We measure components of turbulent velocity fluctuations field in 15 planes as indicated above, and for each velocity field  we determined the average values of $u_i^3$ over the plane. For calculating integral turbulent length scales $\ell_i$  we determine the longitudinal correlation functions of the turbulent velocity field in each plane. These correlation functions are approximated by exponential function and integral scales are calculated by integration of these functions.

It must be noted that the turbulent kinetic energy dissipation rate is different for various components of turbulent velocity fluctuations although the difference is not large. Indeed, the contribution to $\varepsilon$ of the $y$-component is 0.27 while contributions of the turbulent velocity components in the $xz$ plane are 0.35 and 0.38 for $x$  and $z$  components, respectively. When turbulent kinetic energy dissipation rate is calculated through a one-dimensional surrogate $u_y$, as it is done in most experimental investigations, then the magnitude of the constant $C_\varepsilon$  should be multiplied by a factor 1.23 which is the ratio of 1/3 and 0.27. The magnitude of the renormalized constant $C_\varepsilon$ is very close to that obtained in \cite{JC09}. It is conceivable to suggest that one of the reasons of the scatter in the measured value of the constant $C_\varepsilon$  in the cited above studies is caused by anisotropy of the turbulent velocity field.

Using the data obtained from the measured velocity field in many planes we determine the shear-induced production term $P=-\langle u_i \, u_j \rangle \, \nabla_j U_i$ of the turbulent kinetic energy. The main contributions to the production term are due the following terms:
\begin{eqnarray}
P&=&\nu_{_{Tz}} \, \big[(\nabla_z U_y)^2 + (\nabla_z U_x)^2 \big] + \nu_{_{Tx}} \, (\nabla_x U_z)^2
\nonumber\\
&& + \nu_{_{Ty}} \, (\nabla_y U_z)^2 + ... \;,
\label{A21}
\end{eqnarray}
where $\nu_{_{Tx}} = u_x \ell_x$ and similarly for $\nu_{_{Ty}}$ and $\nu_{_{Tz}}$. The production term is averaged over the planes. Other contributions in the production term are small due to a small corresponding large-scale derivative of the mean velocity field or they vanish after integration over $z$  [e.g., $\int_0^{L_z} (\nabla_z U_y) \,dz \approx 0$ if we take into account that the large-scale circulation is nearly anti-symmetric relative to its central line $z=L_z/2$ (see Figs.~\ref{Fig2}-\ref{Fig3}), where $L_z$ is the vertical size of the large-scale circulation].

Our measurements of the turbulent heat flux inside the coherent structures show  that in the regions outside the upward and downward large-scale motions the vertical component of the turbulent heat flux is very small. In particular, the measured vertical turbulent heat flux $F_z=\langle u_i \theta \rangle$ inside the upward and downward large-scale motions for $\Delta T = 60$ K is $F_z = 4$ K cm s$^{-1}$, while outside the upward and downward large-scale motions it is smaller by at least one order of magnitude. For these conditions temperature fluctuations are of the order of $\theta \approx 1.3$ K, the velocity fluctuations are $u_z\approx 5.1$ cm s$^{-1}$ and the correlation coefficient is 0.6. Turbulence production by the vertical turbulent heat flux $P_h = \beta \, F_z$ inside the upward and downward large-scale motions is of the order of $P_h \approx 13.3$ cm$^{2}$ s$^{-1}$  (upper estimate). The area of the vertical flow is of the order of 20 \% (upper estimate) of the area of heat exchangers, which implies that the contribution of the turbulent heat flux to the production turbulence is $0.2 P_h \approx 2.7$ cm$^{2}$ s$^{-1}$.

Measured turbulent energy production due to flow shear, averaged over the central part of the flow, is $P \approx 60.5$ cm$^{2}$ s$^{-1}$ for  $\Delta T = 69$ K. Since in the investigated temperature range the turbulent energy production due to flow shear $P$  is approximately proportional to $\Delta T$, the magnitude of $P$  at $\Delta T = 60$ K  is of the order of  $P \approx 50.2$ cm$^{2}$ s$^{-1}$. Consequently, the contribution of the turbulent heat flux to turbulent energy production can be estimated as 5 \%  (upper estimate).
This implies that the production of the turbulent kinetic energy in the turbulent convection in the regions outside the upward and downward large-scale motions is caused by shearing motions in the coherent structures. In the vicinity of the upward and downward large-scale motions the contribution to production of the turbulent kinetic energy is due to the shear-induced term, $P=-\langle u_i \, u_j \rangle \, \nabla_j U_i$, and to the lesser extent by the buoyancy-induced term, $\beta \, F_z$. Our estimations also show that turbulence inside the large-scale circulations in thermal convection is produced at the expense of a small fraction of thermal energy transported by the flow. In particular, these values are negligibly small (by a factor of  $10^{-5})$  in comparison with a power transported by thermal convection flow from the heater to the cooler.

\begin{figure}
\vspace*{1mm}
\centering
\includegraphics[width=8cm]{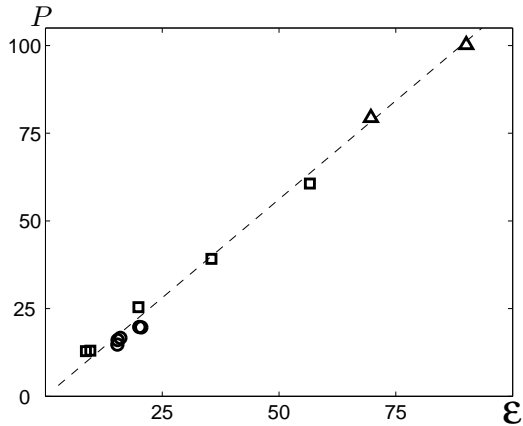}
\caption{\label{Fig5} Dependence of the shear-induced production term $P=-\langle u_i \, u_j \rangle \, \nabla_j U_i$ of the turbulent kinetic energy versus the dissipation rate $\varepsilon$ obtained in the experiments with turbulent convection in the chamber with $A\approx 2$ (triangles for two-cell flow pattern and circles for one-cell flow pattern) and $A\approx 4$ (squares for two-cell flow pattern). The dashed line corresponds to $P=1.1 \varepsilon$. The production rate $P$ and the dissipation rate $\varepsilon$ are measured in cm$^2$ s$^{-3}$.}
\end{figure}

In Fig.~\ref{Fig5} we show the dependence of the shear-induced production term $P=-\langle u_i \, u_j \rangle \, \nabla_j U_i$ of the turbulent kinetic energy versus the dissipation rate $\varepsilon$ in turbulent convection. In our experiments we found that $P=1.1 \varepsilon$, and the data are scattered by 13 \%. This result can be explained using a steady state solution of Eq.~(\ref{A2}) when the production $\beta \, F_z$ caused by the vertical component of the turbulent heat flux is very small. The steady state solution of Eq.~(\ref{A2}) is $P=\varepsilon$. A small deviation from this balance equation observed in the experiments can be explained by that the formula for the dissipation rate of the turbulent kinetic energy includes the empirical constant $C_\varepsilon$, that can deviate from $C_\varepsilon=1$. Another reason for this deviation may be related to the small contribution of the turbulent heat flux to turbulent energy production.

\begin{figure}
\vspace*{1mm}
\centering
\includegraphics[width=8cm]{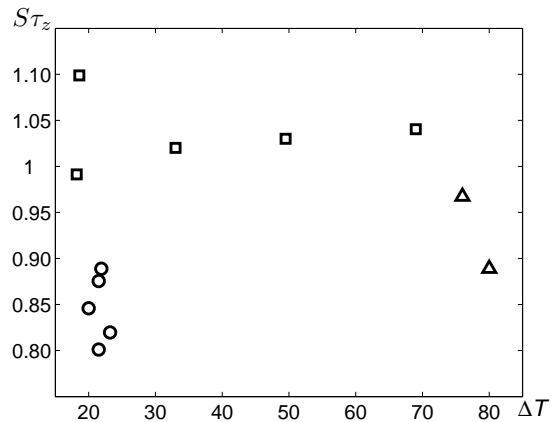}
\caption{\label{Fig6} Measured large-scale shear $S \tau_z$ versus the temperature difference $\Delta T$ between the bottom and the top  walls of the chamber obtained in the experiments with turbulent convection in the chamber with $A\approx 2$ (triangles for two-cell flow pattern and circles for one-cell flow pattern) and $A\approx 4$ (squares for two-cell flow pattern). The temperature difference $\Delta T$ is measured in K.}
\end{figure}

Figure~\ref{Fig6} shows the dependence of the shear $S \tau_z$ versus the temperature difference $\Delta T$ between the bottom and the top walls of the chamber, where $\tau_z= \ell_z / u_z$ is the turbulent time scale along the vertical direction and $S = |\nabla_z U_y|$. Here the parameter $S$ is also averaged over the probed region in the plane $yz$. The obtained dependence is $S \tau_z \approx$ const (in particular, $S \tau_z \approx 1$ for $A\approx 4$ and $S \tau_z \approx 0.9$ for $A\approx 2$). This feature is typical for a shear-produced turbulence (see, e.g., \cite{MY75,KF94}).

\begin{figure}
\vspace*{1mm}
\centering
\includegraphics[width=8cm]{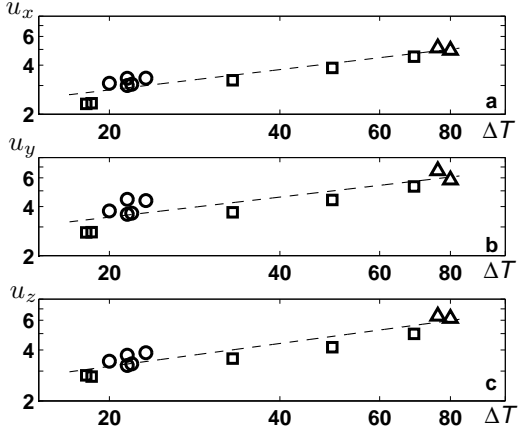}
\caption{\label{Fig7} Component $u_x$ (panel a), $u_y$ (panel b) and $u_z$ (panel c) of the measured turbulent velocity versus the temperature difference $\Delta T$ between the bottom and the top  walls of the chamber (in log-log scale) obtained in the experiments with turbulent convection in the chamber with $A\approx 2$ (triangles for two-cell flow pattern and circles for one-cell flow pattern) and $A\approx 4$ (squares for two-cell flow pattern). The list-square fit for the experimental results is shown by dashed line. The turbulent velocity is measured in cm s$^{-1}$ and the temperature difference $\Delta T$ is measured in K.}
\end{figure}

\begin{figure}
\vspace*{1mm}
\centering
\includegraphics[width=8cm]{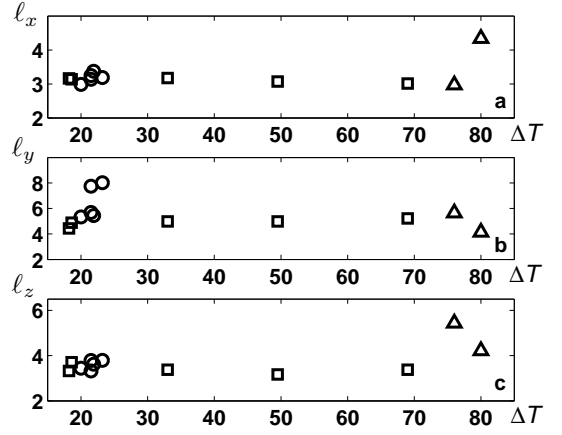}
\caption{\label{Fig8} Turbulent length scales $\ell_x$ (panel a), $\ell_y$ (panel b) and $\ell_z$ (panel c) versus the temperature difference $\Delta T$ between the bottom and the top  walls of the chamber obtained in the experiments with turbulent convection in the chamber with $A\approx 2$ (triangles for two-cell flow pattern and circles for one-cell flow pattern) and $A\approx 4$ (squares for two-cell flow pattern). The turbulent length scales are measured in cm and the temperature difference $\Delta T$ is measured in K.}
\end{figure}

In Fig.~\ref{Fig7} we show the dependence of the components of the measured turbulent velocity $u_x$, $\, u_y$ and $u_z$ versus the temperature difference $\Delta T$ between the bottom and the top  walls of the chamber. The observed scaling for the vertical component of the turbulent velocity is
\begin{eqnarray}
u_z \propto (\Delta T)^{0.45} ,
\label{A3}
\end{eqnarray}
while the observed scaling for the horizontal components $u_x$ and $u_y$ of the turbulent velocity versus $\Delta T$ has a slightly different exponent ($ \approx 0.41$). In Fig.~\ref{Fig8} we show the turbulent length scales $\ell_x$, $\, \ell_y$ and $\ell_z$ versus the temperature difference $\Delta T$. These figures demonstrate that the turbulent length scales are nearly independent of the temperature difference $\Delta T$ between the bottom and the top walls of the chamber.

\begin{figure}
\vspace*{1mm}
\centering
\includegraphics[width=8cm]{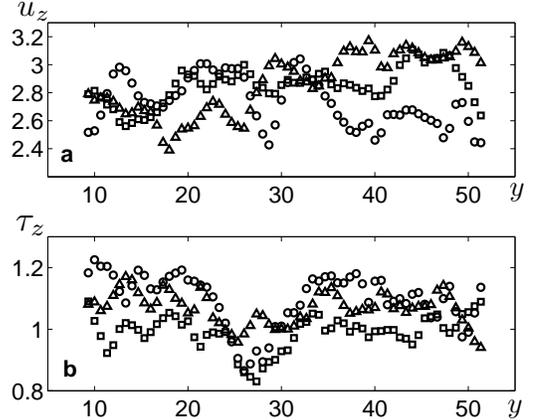}
\caption{\label{Fig9} Dependencies of the vertical component $u_z$ of the measured turbulent velocity (upper panel) and the vertical turbulent time $\tau_z = \ell_z/u_z$ (lower panel) versus the horizontal $y$ coordinate obtained in the experiments with turbulent convection in the chamber with $A\approx 4$ (squares correspond to measurements at the central $yz$ plane at $x=15$ cm, triangles are for the $yz$ plane at $x=5$ cm and circles are for the $yz$ plane at $x=25$ cm). The turbulent velocity is measured in cm s$^{-1}$, the turbulent time is measured in seconds and the coordinate $y$ in cm.}
\end{figure}

It should be also noted that inhomogeneity of the observed turbulent convection is weak, see Fig.~\ref{Fig9} whereby the dependencies of the vertical component $u_z$ of the measured turbulent velocity and the vertical turbulent time $\tau_z = \ell_z/u_z$ versus the horizontal $y$ coordinate are shown.

\begin{figure}
\vspace*{1mm}
\centering
\includegraphics[width=8cm]{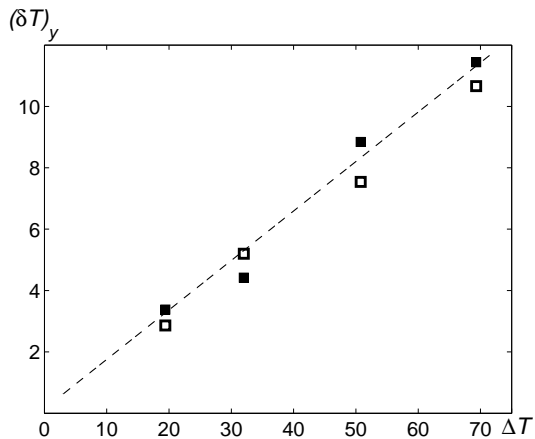}
\caption{\label{Fig10} Dependence of the horizontal temperature difference $(\delta T)_y = |\nabla_y T| L_y$ inside one cell of the two-cell coherent structure versus the vertical temperature difference $\Delta T$ between the bottom and the top  walls of the chamber obtained in the experiments with turbulent convection in the chamber with $A\approx 4$ (the filled squares denote left cell and the unfilled squares denote right cell). Here $L_y$ is the horizontal size in the $y$ direction of the one cell of the coherent structure. The list-square fit for the experimental results is shown by dashed line. The temperature differences $(\delta T)_y$ and $\Delta T$ are measured in K.}
\end{figure}

\begin{figure}
\vspace*{1mm}
\centering
\includegraphics[width=8cm]{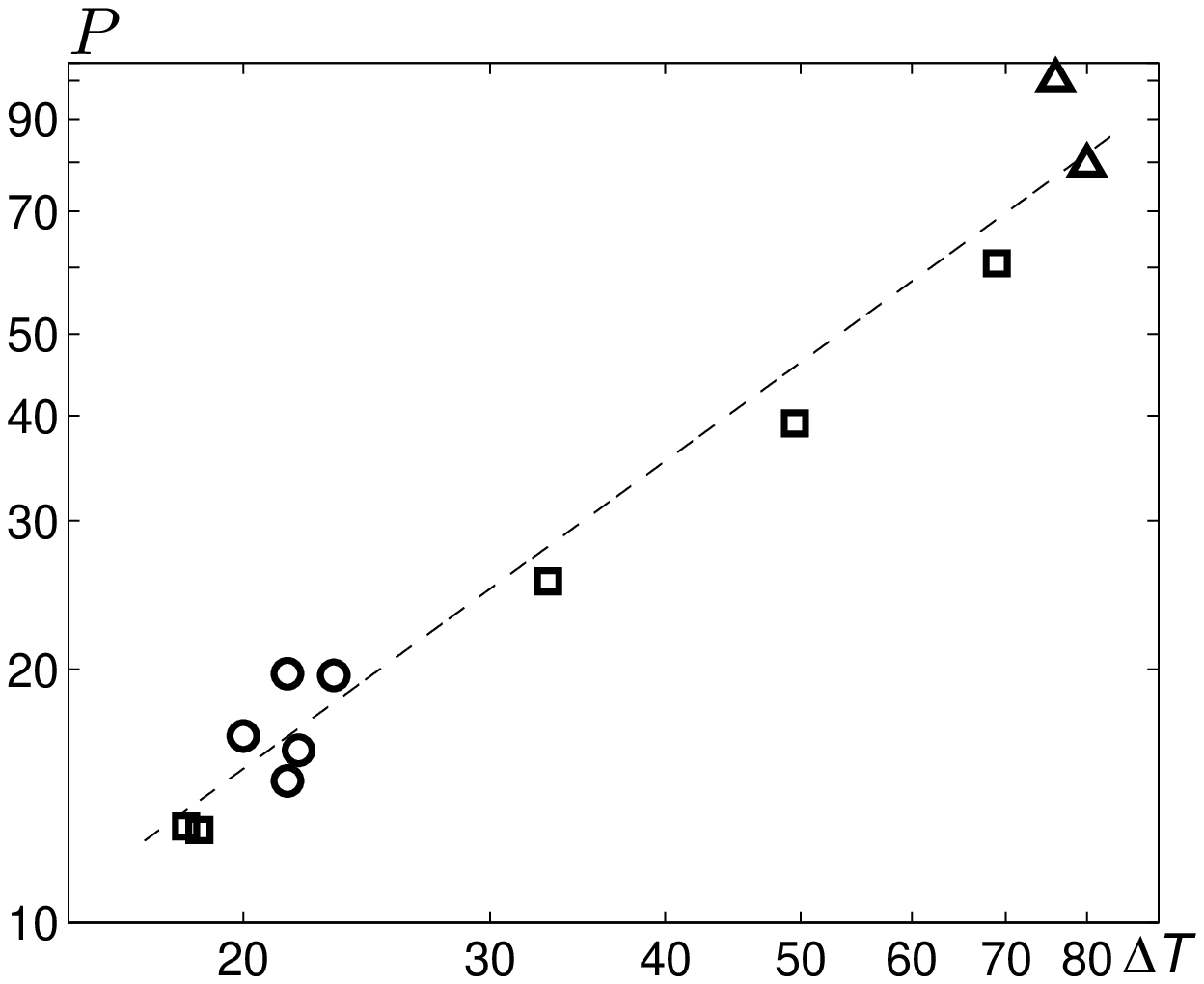}
\caption{\label{Fig11} Production rate $P$ versus the temperature difference $\Delta T$ between the bottom and the top  walls of the chamber (in log-log scale) obtained in the experiments with turbulent convection in the chamber with $A\approx 2$ (triangles for two-cell flow pattern and circles for one-cell flow pattern) and $A\approx 4$ (squares for two-cell flow pattern). The list-square fit for the experimental results is shown by dashed line. The production rate $P$ is measured in cm$^2$ s$^{-3}$ and the temperature difference $\Delta T$ is measured in K.}
\end{figure}

The scaling~(\ref{A3}) can be explained by the following arguments. Consider the $x$ component of {\bf curl} of the momentum equation for the mean velocity ${\bf U}$. In a steady-state this equation yields the following estimate $(g/T_\ast) \, (\nabla_y T) \sim - \nu_{_{T}} \, \Delta (\bec{\nabla} {\bf \times} {\bf U})_x$. Therefore, the velocity shear can be estimated as $S \propto (L_z/\ell)^2 \, (g\tau) \, (\nabla_y T) / T_\ast$, where $S \sim \nabla_z U_y$, $\, T_\ast$ is the characteristic mean temperature inside the coherent structure and $L_z$ is the vertical size of the coherent structure. Here we take into account that $|\nabla_z U_y| \gg |\nabla_y U_z|$, the turbulent viscosity $\nu_{_{T}} \sim u \, \ell = \ell^2 /\tau$ and $- \Delta (\bec{\nabla} {\bf \times} {\bf U})_x \sim S / L_z^2$. On the other hand, our experiments demonstrate that $S \tau =$ const (see, Fig.~\ref{Fig6}). Therefore, the turbulent velocity can be estimated as $u \propto L_z \, (g \, |\nabla_y T| / T_\ast)^{1/2}$.

If we assume that $|\nabla_y T| \sim (\delta T)_y / L_y$  and $(\delta T)_y \propto \Delta T$, we obtain the following scalings for the turbulent velocity field:
\begin{eqnarray}
u \propto (\Delta T)^{1/2} ,
\label{A4}
\end{eqnarray}
and for the dissipation rate of the turbulent kinetic energy:
\begin{eqnarray}
\varepsilon = u^3 / 2 \ell \propto (\Delta T)^{3/2} .
\label{A5}
\end{eqnarray}
In the latter estimate for $\varepsilon$ we assumed that the turbulent length scales $\ell$ are nearly independent of the temperature difference $\Delta T$ between the bottom and the top walls of the chamber (see Fig.~\ref{Fig8}). The assumption about the linear dependence $(\delta T)_y \propto \Delta T$ of the horizontal temperature difference $(\delta T)_y$ on the vertical temperature difference $\Delta T$, is confirmed by our measurements (see Fig.~\ref{Fig10} where the dependence of the horizontal temperature difference $(\delta T)_y = |\nabla_y T| L_y$ inside one cell of the two-cell coherent structure versus the vertical temperature difference $\Delta T$ between the bottom and the top walls of the chamber with $A\approx 4$ obtained in our experiments is shown). Notably, our experiments also show that the shear-induced production of the turbulent kinetic energy scales as
\begin{eqnarray}
P \approx \varepsilon \propto (\Delta T)^{\alpha} ,
\label{A6}
\end{eqnarray}
where $\alpha=1.38$ for $A\approx 4$ and $\alpha=1.21$ for $A\approx 2$ (see Fig.~\ref{Fig11}). A small difference in the exponents in the above estimates and the experimental results might be caused by the weak dependence of the turbulent length scales on the temperature difference $\Delta T$, and a small deviation from the linear dependence $(\delta T)_y \propto \Delta T$.

In this study we presented the dependencies of the turbulent flow parameters versus the temperature difference $\Delta T$ between the bottom and the top walls of the chamber rather than on global Rayleigh number, Ra =$\beta \, (\Delta T) \, H_z^3 /(\nu \, \kappa)$, because of the following reasons. Our experiments among the others demonstrate that the life-time of large-scale circulations (LSC) is much larger than the characteristic turbulence time scale $\tau$, and characteristic spatial scales of LSC are much larger than the characteristic turbulence spatial scales. The time of transport of a fluid particle through LSC is much larger than the time $\tau$. This implies that LSC can be described in terms of a mean-field approach, i.e., LSC is a mean-field object. Therefore, turbulent viscosity and turbulent temperature diffusivity, which are much larger than molecular transport coefficients (the kinematic viscosity, $\nu$, and the temperature diffusivity, $\kappa$), become more important. Consequently,  the effective Rayleigh number based on turbulent viscosity and turbulent temperature diffusivity, is an important parameter characterizing the properties of LSC. It is plausible to suggest also that when the global Rayleigh number is much larger than the threshold value required for the excitation of convection, the properties of LSC are weakly dependent on the magnitude of the global Rayleigh number.

\begin{figure}
\vspace*{1mm}
\centering
\includegraphics[width=8cm]{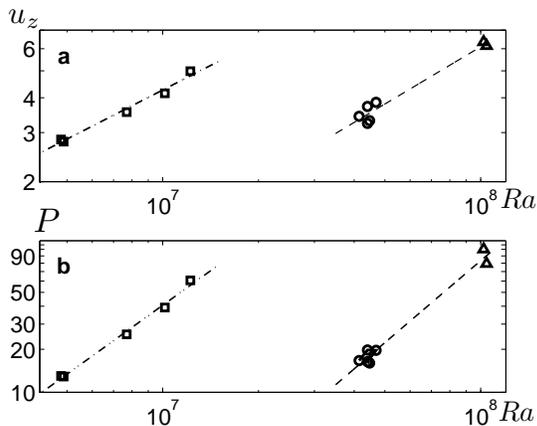}
\caption{\label{Fig12} Dependencies of the vertical component $u_z$ of the measured turbulent velocity (upper panel) and the production rate $P$ (lower panel) versus the global Rayleigh number ${\rm Ra}$ (in log-log scale) obtained in the experiments with turbulent convection in the chamber with $A\approx 2$ (triangles for two-cell flow pattern and circles for one-cell flow pattern) and $A\approx 4$ (squares for two-cell flow pattern). The list-square fit for the experimental results is shown by dashed line for $A\approx 2$ and by dashed-dotted line for $A\approx 4$. The production rate $P$ is measured in cm$^2$ s$^{-3}$ and the turbulent time is measured in cm s$^{-1}$.}
\end{figure}

We show the dependencies of the turbulent flow parameters versus the temperature difference $\Delta T$ between the bottom and the top walls of the chamber with two aspect ratios $A\approx 2$ and $A\approx 4$ in the same figure. When the temperature difference $\Delta T$ is varied, the kinematic viscosity $\nu(T)$, the temperature diffusivity $\kappa(T)$ and the parameter $\beta(T)$ which depend on the mean fluid temperature, are also changed. Therefore, several variables $(\Delta T$, $\nu(T)$, $\kappa(T)$  and $\beta(T)$) in the expression for the global Rayleigh number are changed when the temperature difference $\Delta T$ is varied. Another reason for using the temperature difference $\Delta T$ rather than the global Rayleigh number as independent variable is that we determined  theoretically the scalings of the production rate $P$ and turbulent (rms) velocity ${\bf u}$ with the temperature difference $\Delta T$ rather than on global Rayleigh number, Ra. However, for comparing our results with other studies we also presented some dependencies of the turbulent flow parameters versus global Rayleigh number (see Fig.~\ref{Fig12}).

\section{Discussion and conclusions}

We investigated the effect of large-scale coherent structures on global properties of turbulent convection in laboratory experiments. The turbulent convection for the Rayleigh numbers varying from $5 \times 10^6$ to $10^8$, was studied  in air flow in a rectangular chamber with aspect ratios $A\approx 2$ and $A\approx 4$. The large-scale coherent structures comprise the one-cell and two-cell flow patterns. It was demonstrated that turbulent convection with large-scale coherent structures is a shear-produced turbulence. We determined the dependence of global parameters (the production and dissipation of turbulent kinetic energy, the turbulent velocity and integral turbulent scale, the large-scale shear, etc.) of turbulent convection on the temperature difference between the bottom and the top walls of the chamber. These dependencies are in an agreement with theoretical predictions.

Equation~(\ref{A1}) that is valid for a a buoyancy-produced turbulence, yields much smaller values for the turbulent kinetic energy than those observed in our experiments. The production of the turbulent kinetic energy in turbulent convection, caused by the buoyancy flux, even in the upward and downward flow regions is much smaller than the production rate due to the shear motions inside the large-scale coherent structures. Inside the coherent structures where the vertical heat flux is very small, the production of the turbulent kinetic energy, caused by the buoyancy flux, is by several orders of magnitude smaller than the production rate due to the shear motions.

The large-scale coherent structures are formed due to large-scale instability caused by a redistribution of the turbulent heat flux due to non-uniform motions. As soon as the large-scale coherent structures are formed, there is a strong influence of the shear produced by the coherent structures. In particular, the main contribution to production of turbulence is due to shearing motions inside the large-scale coherent structures. The magnitude of the large-scale shear is determined by the horizontal mean temperature gradient inside the large-scale circulation. These features are different from those which are typical for the classical laminar convection.

\begin{acknowledgements}
We thank A.~Krein for his assistance in construction of the
experimental set-up. This research was supported in part by the Israel Science Foundation governed by the Israeli Academy of Science.
\end{acknowledgements}

\end{document}